\documentclass{INTERSPEECH2023}
\usepackage{multirow}
\usepackage{makecell}

\interspeechcameraready

\title{The DKU-MSXF Diarization System for the VoxCeleb Speaker Recognition Challenge 2023}
\name{Ming~Cheng$^1$, Weiqing~Wang$^1$, Xiaoyi~Qin$^1$, Yuke~Lin$^1$, Ning~Jiang$^2$, Guoqing~Zhao$^2$, Ming~Li$^1$}
\address{
  $^1$Data Science Research Center, Duke Kunshan University, Kunshan, China \\
  $^2$Mashang Consumer Finance Co., Ltd.}
\email{ming.li369@dukekunshan.edu.cn}

\begin{document}

\maketitle
 
\begin{abstract}
This paper describes the DKU-MSXF submission to track 4 of the VoxCeleb Speaker Recognition Challenge 2023 (VoxSRC-23). Our system pipeline contains voice activity detection, clustering-based diarization, overlapped speech detection, and target-speaker voice activity detection, where each procedure has a fused output from 3 sub-models. Finally, we fuse different clustering-based and TSVAD-based diarization systems using DOVER-Lap and achieve the 4.30\% diarization error rate (DER), which ranks first place on track 4 of the challenge leaderboard.

\end{abstract}

\section{Introduction}

Speaker diarization is the task of breaking up multi-party conversational audio into speaker-homogeneous segments, which aims to solve the problem of "Who-Spoke-When." In this paper, we focus on the speaker diarization task and present the details of our submitted system to track 4 of the VoxCeleb Speaker Recognition Challenge 2023.

\begin{figure}[t]
\centering
  \includegraphics[width=\linewidth]{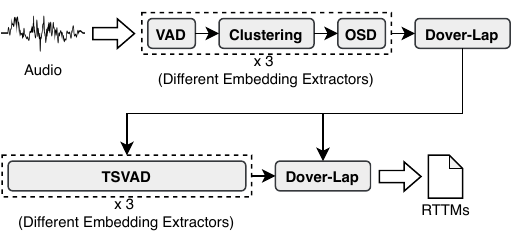}
  \caption{Framework of The Developed System.}
  \label{fig:framework}
\end{figure}

Fig.~\ref{fig:framework} depicts the framework of our developed system. First, the voice activity detection (VAD) module removes non-speech regions from the input audio. The remaining is split into multiple short segments, followed by speaker embedding extraction. Then, the initial clustering-based diarization results can be obtained by agglomerative hierarchical clustering (AHC) with overlapped speech detection (OSD) as the post-processing. We replace speaker embedding models trained under different conditions to repeat the above process three times and obtain the fused clustering-based results by Dover-Lap~\cite{raj2021dover}. Next, target-speaker voice activity detection (TSVAD) models with different speaker embedding models are adopted to refine the clustering-based results. In the end, clustering-based and TSVAD-based results are fused again to obtain the final prediction.

In general, the framework integrates the advantages of both clustering-based and TSVAD-based diarization methods, which is similar to our previous submissions in VoxSRC-21~\cite{wang2021dkudukeecelenovo} and VoxSRC-22~\cite{wang2022dkudukeece}. The main differences are improved speaker embedding models and Seq2Seq-TSVAD models from our recent works~\cite{cheng2023target,cheng2023whu}.

\section{Dataset Description}

According to challenge rules, any data except the challenge test data is allowed in this task. The datasets used to train each model in our system are described as follows.

\begin{itemize}
	\item Voice activity detection (VAD) and overlapped speech detection (OSD): VoxCeleb 1\&2~\cite{Nagrani17,Chung18b} for data simulation and VoxConverse~\cite{chung2020spot} for adaptation and validation.
	\item Speaker embedding: VoxCeleb 1\&2~\cite{Nagrani17,Chung18b} and VoxBlink-Clean~\cite{lin2023voxblink} for training and evaluation.
	\item Clustering-based diarization: VoxConverse~\cite{chung2020spot} for hyper-parameter tuning.
	\item TSVAD-based diarization: VoxCeleb 1\&2~\cite{Nagrani17,Chung18b} for online data simulation and VoxConverse~\cite{chung2020spot} for adaptation and validation. 
	\item Data augmentation: MUSAN~\cite{snyder2015musan} and RIRs\cite{ko2017study} corpora.
\end{itemize}

\section{Model Configuration}

This section describes each model in our submitted system. If not specified, the input acoustic features of all models are 80-dim log Mel-filterbank energies with a frame length of 25 ms and a frameshift of 10 ms. MUSAN~\cite{snyder2015musan} and RIRs~\cite{ko2017study} are applied as the data augmentation.

\subsection{VAD \& OSD}

As the implementations of voice activity detection (VAD) and overlapped speech detection (OSD) tasks are very similar, we utilize the same neural networks to train these two tasks, respectively. Adopted model architectures are described as follows.

\subsubsection{Conformer}

Conformer~\cite{gulati2020conformer} is the first backbone network. All encoder layers share the same settings: 256-dim attentions with 4 heads and 1024-dim feed-forward layers with a dropout rate of 0.1. The kernel size of convolutions in Conformer blocks is 15. Finally, a linear layer with sigmoid activation is adopted to transform the dimension of Conformer outputs into one, representing the frame-level posterior probability of VAD or OSD.

\subsubsection{ResNet34}

ResNet34~\cite{he2016deep} is the second backbone network, where the widths (number of channels) of the residual blocks are $\left \{64, 128, 256, 512 \right \}$. At the end of convolutional blocks, the spatial average pooling extracts frame-level features from the convolutional outputs. Finally, a linear layer with sigmoid activation predicts the frame-level posterior probability of VAD or OSD.

\subsubsection{ECAPA-TDNN}

ECAPA-TDNN~\cite{desplanques20_interspeech} is the third backbone network. The number of filters in the convolutional layers is set to 1024. The scale dimension in the Res2Block is set to 8. The dimension of the bottleneck in the SE-Block and the attention module is set to 128. Finally, a linear layer with sigmoid activation predicts the frame-level posterior probability of VAD or OSD.

\begin{table}[t]
	\centering
	\renewcommand{\arraystretch}{1.1}
	\caption{False alarm (FA) and miss detection (MISS) rates of different VAD and OSD models on the VoxConverse test set.}
	\label{tab:vad_and_osd}
	\begin{tabular}{llrrr}
		\toprule
		 \textbf{Task} &\textbf{Model} & \textbf{FA (\%)} & \textbf{MI (\%)} & \textbf{Total (\%)}\\
		\midrule
		\multirow{4}{*}{VAD}
		& Conformer  & 2.84 & 1.09 & 3.93 \\
		& ResNet34   & 3.20 & 1.02 & 4.22 \\
		& ECAPA-TDNN & 2.70 & 1.51 & 4.21 \\
		& Fusion     & 2.83 & 1.14 & 3.97 \\
		\midrule
		\multirow{4}{*}{OSD}
		& Conformer  & 0.59 & 1.41 & 2.00 \\
		& ResNet34   & 0.54 & 1.51 & 2.05 \\
		& ECAPA-TDNN & 0.52 & 1.45 & 1.97 \\
		& Fusion     & 0.44 & 1.45 & 1.89 \\
		\bottomrule
	\end{tabular}
\end{table}

\subsubsection{Fusion}

VAD or OSD models with different network architectures are fused by averaging their predictions at the score level. Table~\ref{tab:vad_and_osd} shows the VAD and OSD models' performance on the VoxConverse test set, respectively. We adopt the fused results as the final predictions in this part. 

\subsection{Speaker Embedding}

To the goal of ensemble learning, we train three speaker embedding models with diverse network architectures and training data. The first one is SimAM-ResNet34~\cite{qin2022simple} with statistics pooling (SP)~\cite{snyder2018x}. The second one is ResNet101~\cite{he2016deep} with attentive statistics pooling (ASP)~\cite{okabe18_interspeech}. The third one is SimAM-ResNet100 with ASP. The first two models are trained on VoxCeleb2~\cite{Chung18b} dataset. For the last model, we additionally mix the VoxBlink-Clean~\cite{lin2023voxblink} dataset into the training.

\begin{table}[t]
	\centering
	\setlength{\tabcolsep}{3.2pt}
	\renewcommand{\arraystretch}{1.1}
	\caption{Equal error rates (EERs) of different speaker embedding models on the Vox-O trial.}
	\label{tab:spk_embd}
	\begin{tabular}{lllr}
		\toprule
		\textbf{\#} & \textbf{Model} & \textbf{Training Data} & \textbf{EER (\%)} \\
		\midrule
		Spk1 & SimAM-ResNet34+SP   & Vox2 & 0.81 \\
		Spk2 & ResNet101+ASP       & Vox2 & 0.49 \\
		Spk3 & SimAM-ResNet100+ASP & Vox2+VoxBlink & 0.44 \\
		\bottomrule
	\end{tabular}
\end{table}

All speaker embedding models utilize the same back-end part. After the pooling layer projects the variable-length input audio to the fixed-length vector, a 256-dim fully connected layer is adopted as the speaker embedding layer, and the ArcFace (s=32,m=0.2)~\cite{deng2019arcface} is used as a classifier. The detailed configuration of model training is the same as~\cite{qin2022simple}. The performance of different trained speaker embedding models is shown in Table~\ref{tab:spk_embd}.

\subsection{Clustering-based Diarization}

We adopt agglomerative hierarchical clustering (AHC) to implement the clustering-based diarization system, which is the same as we used in previous years~\cite{wang2021dkudukeecelenovo,wang2022dkudukeece}. First, speaker embeddings are extracted from the uniformly segmented speech with a length of 1.28s and a shift of 0.32s, pairwisely measured by cosine similarity. Two consecutive segments are merged into a longer segment if their similarity exceeds a segment threshold. Then, we perform a plain AHC on the similarity matrix with a relatively high stop threshold to obtain clusters with high confidence. These clusters are split into "long clusters" and "short clusters" by the total duration in each cluster, and the central embedding of each cluster is the mean of all speaker embeddings within the cluster. Finally, each short cluster is assigned to the closest long cluster by the similarity between their central embeddings. If a short cluster is too different from all long clusters, the similarity between them is lower than a speaker threshold, and then we treat it as a new speaker. As post-processing based on the OSD model, overlapped speech regions are assigned to the two closest speaker labels.

Based on speaker embedding models in Table~\ref{tab:spk_embd}, we develop three AHC-based diarization systems following the above approach. All the hyper-parameters are directly tuned on the VoxConverse test set by grid search. The duration for classifying long and short clusters is 6s for all models. The other thresholds are shown in Table~\ref{tab:ahc}.

\begin{table}[t]
	\centering
	\setlength{\tabcolsep}{7.85pt}
	\renewcommand{\arraystretch}{1.1}
	\caption{Thresholds (THRs) of different AHC-based diarization systems.}
	\label{tab:ahc}
	\begin{tabular}{lrrrr}
		\toprule
		\textbf{\#} & \textbf{Segment THR} & \textbf{Stop THR} & \textbf{Speaker THR}\\
		\midrule
		Ahc1 & 0.54 & 0.60 & 0.20 \\
		Ahc2 & 0.62 & 0.62 & 0.20 \\ 
		Ahc3 & 0.66 & 0.68 & 0.30 \\
		\bottomrule
	\end{tabular}
\end{table}

\subsection{TSVAD-based Diarization} 

Given speaker profiles (e.g., x-vector~\cite{snyder2018x}), the TSVAD method can estimate each speaker's frame-level voice activities and perform robustly even in complex acoustic environments. We adopt the Seq2Seq-TSVAD~\cite{cheng2023target} consisting of the front-end extractor with segmental pooling~\cite{9849033}, Conformer~\cite{gulati2020conformer} encoder, and proposed speaker-wise decoder. Specifically, the front-end extractor is initialized by the pre-trained speaker embedding model same as the one for extracting speaker profiles.

Each model training starts from BCE loss \& Adam optimizer with a learning rate of $1e-4$ and a linear warm-up of 2,000 iterations. The whole process is described as follows.

\begin{table*}[t]
	\centering
	\setlength{\tabcolsep}{12pt}
	\renewcommand{\arraystretch}{1.1}
	\caption{Diarization error rates (DERs) of different systems on the VoxConverse test set, VoxConverse test46 set, and VoxSRC-23 challenge test set.}
	\label{tab:exps}
	\begin{tabular}{llrrr}
		\toprule
		\multirow{2}{*}{\textbf{\#}} &
		\multirow{2}{*}{\textbf{Method}} & \multicolumn{3}{c}{\textbf{DER (\%)}} \\
		\cmidrule(lr){3-5}
		& & \textbf{VoxConverse Test} & \textbf{VoxConverse Test46} & \textbf{VoxSRC-23 Challenge Test}  \\		
		\midrule
		1 & Ahc1                & 4.83 & 4.14 & 5.51    \\
		2 & Ahc2                & 4.49 & 3.92 & 5.32    \\
		3 & Ahc3                & 4.55 & 3.92 & 5.36    \\
		\midrule
	    4 & Dover-Lap (\#1-3)         & - & 3.81 & 5.19 \\
	    5 & \quad $+$ TSVAD with Spk1 & - & 2.85 & 4.49 \\
	    6 & \quad $+$ TSVAD with Spk2 & - & 2.93 & 4.57 \\
	    7 & \quad $+$ TSVAD with Spk3 & - & 2.91 & 4.53 \\
	    \midrule
	    8 & Dover-Lap (\#4-7)         & - & 2.73 & 4.30 \\
		\bottomrule
	\end{tabular}
\end{table*}

\begin{itemize}
  \item First, the model with a frozen front-end extractor is trained on simulated data until back-end convergence.
  \item Second, all model parameters are unfrozen to train on both simulated (VoxCeleb 1\&2) and real data (VoxConverse dev set) at the ratio of $0.8/0.2$. Also, the learning rate decreases to $1e-5$ in the second half of this phase.
  \item Third, all data simulation and augmentation are removed, inspired by the large margin finetuning (LMFT)~\cite{9414600} in speaker verification. Meanwhile, to use much training data as possible, we mix the first 186 samples of the VoxConverse test set into finetuning and leave the last 46 samples as validation, namely the VoxConverse test46 set.
\end{itemize}

During inference, a clustering-based diarization is required first to extract speaker profiles from each speaker's speech segments. Then, each test audio is cut into fixed-length chunks with a stride of 1s and fed into the TSVAD model with extracted speaker profiles. Chunked predictions are stitched by averaging the overlapped predicted regions, which can also be viewed as a score-level fusion. 

Using speaker embedding models in Table~\ref{tab:spk_embd} as different profile extractors, we develop three TSVAD-based diarization systems. The first one equipped with SimAM-ResNet34 is trained and inferred under audio chunks of 64s. The last two equipped with ResNet101 and SimAM-ResNet100 are trained and inferred under audio chunks of 16s, limited by higher GPU memory costs of the deeper networks. The other configurations share the same settings: 512-dim attentions with 8 heads, convolutions with a kernel size of 15, and 1024-dim feed-forward layers with a dropout rate of 0.1. The speaker capacity and output VAD resolution are set to 30 and 0.08s, respectively. 

\section{Experimental Results}

Table~\ref{tab:exps} illustrates the performance of our developed diarization systems on the VoxConverse test set, VoxConverse test46 set, and VoxSRC-23 challenge test set. As the TSVAD models utilize part of the VoxConverse test set in training, their performance is only evaluated on the other two datasets.

Systems \#1-3 represent the AHC-based diarization with different speaker embedding models, obtaining the DERs of 5.51\%, 5.32\%, and 5.36\% on the challenge test set. Then, we adopt the Dover-Lap method to fuse them and obtain system \#4 to achieve a 5.19\% DER on the challenge test set.

Systems \#5-7 represent the TSVAD-based diarization with different speaker embedding models. Using the fused clustering-based results to extract speaker profiles, these three models have very close DERs varying from 4.49\% to 4.57\% on the challenge test set. Finally, systems \#4-7 are fused by the Dover-Lap again to obtain system \#8, which achieves the best 4.30\% DER on the challenge test set.

\section{Conclusions}

This paper describes our system development for track 4 of the VoxSRC-23. This year, we mainly focus on improving the speaker embedding, AHC-based diarization, and TSVAD-based diarization models. To achieve the best performance, we also train diverse sub-models for each part of the whole framework. The finally fused method shows significant improvement, obtaining the DERs of 2.73\% on the VoxConverse test46 set and 4.30\% on the VoxSRC-23 challenge test set, respectively.

\bibliographystyle{IEEEtran}
\bibliography{refs}

\end{document}